\documentstyle[amssymb,epsfig]{new}
\begin{document}
%
%
\title{Four-Neutrino Mixing, Oscillations and Big-Bang Nucleosynthesis}
\author{S.M. Bilenky\\
{\it Joint Institute for Nuclear Research, Dubna, Russia, and
Institut f\"ur Theoretische Physik,
Technische Universit\"at Munchen, D--85748 Garching, Germany}\\
C. Giunti\\
{\it INFN, Sezione di Torino, and Dipartimento di Fisica Teorica,
Universit\`a di Torino,
Via P. Giuria 1, I--10125 Torino, Italy}\\
W. Grimus and T. Schwetz\\
{\it Institute for Theoretical Physics, University of Vienna,
Boltzmanngasse 5, A--1090 Vienna, Austria}}

\maketitle

\section*{Abstract}

We investigate the implications
of the standard Big-Bang Nucleosynthesis
constraint on the number of light neutrinos
in the framework of the two four-neutrino schemes
that are favored by the results
of neutrino oscillation experiments.\\
Talk presented by C. Giunti at the Symposium
"New Era in Neutrino Physics",
Tokyo Metropolitan University,
Japan, 11-12 June, 1998.\\
DFTT 51/98, TUM-HEP-328/98, SFB 375-308.

\section{Introduction}
\label{Introduction}

Forty years after Pontecorvo's discovery
of the theory of neutrino oscillations~\cite{Pontecorvo},
there are today three experimental indications
in favor of the existence of this phenomenon.
The strongest evidence is believed to come from
the atmospheric neutrino anomaly
observed by the Super-Kamiokande experiment~\cite{SK-atm},
following the previous measurements of the
Kamiokande,
IMB
and
Soudan
atmospheric neutrino experiments
\cite{Kam-atm-IMB-Soudan}.
The atmospheric neutrino anomaly consists in a suppression of
the ratio of measured atmospheric $\nu_\mu$ and $\nu_e$
with respect to the calculated one
(see, for example, Ref.~\cite{Gaisser}).
Furthermore,
the new data of the Super-Kamiokande experiment~\cite{SK-atm}
indicate a variation of the atmospheric $\nu_\mu$ flux
as a function of the zenith angle,
which correspond to a variation as a function of the distance
between the
region in the atmosphere where the neutrino is produced
and the Super-Kamiokande detector where the neutrino is detected.
This is a strong evidence in favor of neutrino oscillations
and seems to be supported by the measurement of the
flux of neutrino-induced upward-going muons
of the MACRO experiment~\cite{MACRO}.

The oldest indication in favor of neutrino oscillations
comes from the solar neutrino problem
(see, for example, Ref.~\cite{BKS98}),
\textit{i.e.}
the suppression of the flux of solar $\nu_e$'s
observed in five solar neutrino experiments
(Homestake,
Kamiokande,
GALLEX,
SAGE
and
Super-Kamiokande
\cite{solar-nu-exp})
with respect to the calculated one~\cite{BP98}.

Finally,
the LSND accelerator experiment
has reported evidence in favor of
$\bar\nu_\mu\to\bar\nu_e$
and
$\nu_\mu\to\nu_e$ oscillations~\cite{LSND}.

Neutrino oscillations
can occur only if neutrinos are massive particles,
if their masses are different
and if neutrino mixing is realized in nature.
In this case,
the left-handed flavor neutrino
fields
$\nu_{{\alpha}L}$
($\alpha=e,\mu,\tau$)
are superpositions
of
the left-handed components
$\nu_{kL}$
($k=1,\ldots,n$)
of the fields of neutrinos with definite masses
$m_k$:
$
\nu_{{\alpha}L}
=
\sum_{k=1}^{n}
U_{{\alpha}k}
\,
\nu_{kL}
\,,
$
where $U$
is a unitary mixing matrix.
The general expression for the probability of
$\nu_\alpha\to\nu_\beta$
transitions in vacuum is
\begin{equation}
P_{\nu_\alpha\to\nu_\beta}
=
\left|
\sum_{k=1}^{n}
U_{{\beta}k}
\,
\exp\left( - i \, \frac{ \Delta{m}^2_{k1} \, L }{ 2 \, p } \right)
\,
U_{{\alpha}k}^*
\right|^2
\,,
\label{Posc}
\end{equation}
where
$ \Delta{m}^2_{kj} \equiv m_k^2-m_j^2 $,
$L$ is the distance between the neutrino source and detector
and $p$ is the neutrino momentum.

\section{Four neutrino mixing and oscillations}
\label{Four neutrino mixing and oscillations}

The three experimental indications
in favor of neutrino oscillations
require the existence of
three different scales
of neutrino mass-squared differences:
$
\Delta{m}^2_{{\rm sun}}
\sim
10^{-5} \, {\rm eV}^2
$
(MSW)
or
$
\Delta{m}^2_{{\rm sun}}
\sim
10^{-10} \, {\rm eV}^2
$
(vacuum oscillations),
$
\Delta{m}^2_{{\rm atm}}
\sim
5 \times 10^{-3} \, {\rm eV}^2
$
and
$
\Delta{m}^2_{{\rm SBL}} \sim 1 \, {\rm eV}^2
$,
where
$\Delta{m}^2_{{\rm SBL}}$
is the neutrino mass-squared difference
relevant for short-baseline (SBL) experiments,
whose allowed range is determined by the
positive result of the LSND experiment.

Three independent mass-squared differences
require at least four massive neutrinos.
The number of active light flavor neutrinos is known to be three
from the measurement of the invisible width of the $Z$-boson,
but there is no experimental upper bound for
the number of massive neutrinos
(the lower bound is three).
In the following we consider the simplest possibility
of existence of four massive neutrinos
\cite{four,OY96,BGKP,BGG96,BGG97a,BGG97b,BGGS98}.
In this case,
in the flavor basis,
besides the three light flavor neutrinos
$\nu_e$,
$\nu_\mu$,
$\nu_\tau$
that
contribute to the invisible width of the $Z$-boson,
there is a light sterile neutrino
$\nu_s$
that is a SU$(2)_L$ singlet
and does not take part in
standard weak interactions.

Two years ago we have shown~\cite{BGG96}
that among all the possible
four-neutrino mass spectra
only two are compatible
with the results of all neutrino oscillation experiments:
\begin{equation}\label{spectrum}
\mbox{(A)}
\qquad
\underbrace{
\overbrace{m_1 < m_2}^{{\rm atm}}
\ll
\overbrace{m_3 < m_4}^{{\rm sun}}
}_{{\rm SBL}}
\qquad \mbox{and} \qquad
\mbox{(B)}
\qquad
\underbrace{
\overbrace{m_1 < m_2}^{{\rm sun}}
\ll
\overbrace{m_3 < m_4}^{{\rm atm}}
}_{{\rm SBL}}
\,.
\label{AB}
\end{equation}
In these two schemes
the four neutrino masses
are divided in two pairs of close masses
separated by a gap of about 1 eV,
which provides the mass-squared difference
$ \Delta{m}^2_{{\rm SBL}} = \Delta{m}^2_{41} \equiv m_4^2 - m_1^2 $
that is relevant for the oscillations
observed in the LSND experiment.
In scheme A
$ \Delta{m}^2_{{\rm atm}} = \Delta{m}^{2}_{21} \equiv m_2^2 - m_1^2 $
is relevant
for the explanation of the atmospheric neutrino anomaly
and
$ \Delta{m}^2_{{\rm sun}} = \Delta{m}^{2}_{43} \equiv m_4^2 - m_3^2 $
is relevant
for the suppression of solar $\nu_e$'s,
whereas in scheme B
$ \Delta{m}^2_{{\rm atm}} = \Delta{m}^{2}_{43} $
and
$ \Delta{m}^2_{{\rm sun}} = \Delta{m}^{2}_{21} $.

Let us define the quantities $c_\alpha$,
with $\alpha=e,\mu,\tau,s$,
in the schemes A and B as
\begin{equation}
({\rm A})
\quad
c_\alpha
\equiv
\sum_{k=1,2}
|U_{{\alpha}k}|^2
\,,
\qquad
({\rm B})
\quad
c_\alpha
\equiv
\sum_{k=3,4}
|U_{{\alpha}k}|^2
\,.
\label{04}
\end{equation}
Physically $c_\alpha$ quantify the mixing of the flavor neutrino $\nu_\alpha$
with the two massive neutrinos whose $\Delta{m}^2$
is relevant for the oscillations of atmospheric neutrinos
($\nu_1$, $\nu_2$ in scheme A
and
$\nu_3$, $\nu_4$ in scheme B).

The probability of
$\nu_\alpha\to\nu_\beta$
transitions ($\beta\neq\alpha$)
and the survival probability of $\nu_\alpha$
in SBL experiments
are given by~\cite{BGKP}
\begin{equation}
P_{\nu_\alpha\to\nu_\beta}
=
A_{\alpha;\beta} \, \sin^2 \frac{ \Delta{m}^2_{{\rm SBL}} L }{ 4 p }
\,,
\qquad
P_{\nu_\alpha\to\nu_\alpha}
=
1
-
B_{\alpha;\alpha} \, \sin^2 \frac{ \Delta{m}^2_{{\rm SBL}} L }{ 4 p }
\,,
\label{prob}
\end{equation}
with
the oscillation amplitudes
\begin{equation}
A_{\alpha;\beta}
=
4
\left|
\sum_k
U_{{\beta}k}
\,
U_{{\alpha}k}^*
\right|^2
\,,
\qquad
B_{\alpha;\alpha}
=
4
c_\alpha \, ( 1 - c_\alpha )
\,,
\label{ampli}
\end{equation}
where the index $k$ runs over the values $1,2$ or $3,4$.
The probabilities (\ref{prob})
have the same form as the corresponding probabilities in the
case of two-neutrino mixing,
$
P_{\nu_\alpha\to\nu_\beta}
=
\sin^2(2\theta) \, \sin^2(\Delta{m}^2L/4p)
$
and
$
P_{\nu_\alpha\to\nu_\alpha}
=
1
-
\sin^2(2\theta) \, \sin^2(\Delta{m}^2L/4p)
$,
which have been used by all experimental collaborations
for the analysis of the data in order to get information
on the parameters
$\sin^2(2\theta)$
and
$\Delta{m}^2$
($\theta$ and $\Delta{m}^2$ are, respectively, the mixing angle
and the mass-squared difference in the case of two-neutrino mixing).
Therefore,
we can use the results of their analyses in order to get information
on the corresponding parameters
$A_{\alpha;\beta}$,
$B_{\alpha;\alpha}$
and
$\Delta{m}^2_{{\rm SBL}}$.

The results of all neutrino oscillation experiments
are compatible with the schemes A and B only
if~\cite{BGG96}
\begin{equation}
c_e \leq a_e^0
\quad \mbox{and} \quad
c_\mu \geq 1 - a_\mu^0
\,,
\label{03}
\end{equation}
where
\begin{equation}
a_\alpha^0
\equiv
\frac{1}{2}
\left( 1 - \sqrt{ 1 - B_{\alpha;\alpha}^0 } \, \right)
\qquad
(\alpha=e,\mu)
\label{a0}
\end{equation}
and
$B_{\alpha;\alpha}^0$
is the upper bound for the amplitude of
$
\stackrel{\scriptscriptstyle(-)}{\nu}_{\hskip-3pt\alpha}
\to
\stackrel{\scriptscriptstyle(-)}{\nu}_{\hskip-3pt\alpha}
$
oscillations
obtained from the exclusion plots of
SBL reactor and accelerator disappearance experiments.
Hence,
the quantities
$a_e^0$ and $a_\mu^0$
depend on $\Delta{m}^2$.
The exclusion curves obtained in the Bugey reactor experiment~\cite{Bugey95}
and in the CDHS and CCFR accelerator
experiments~\cite{CDHS84-CCFR84}
imply that both
$a_e^0$ and $a_\mu^0$
are small~\cite{BBGK96}:
$ a^{0}_e \lesssim 4 \times 10^{-2} $
and
$ a^{0}_\mu \lesssim 2 \times 10^{-1} $
for any value of
$\Delta{m}^{2}$
in the range
$
0.3
\lesssim
\Delta{m}^2
\lesssim
10^{3} \, {\rm eV}^2
$.

The smallness of $c_e$
in both schemes A and B
is a consequence of the solar neutrino problem~\cite{BGG96}.
It implies that the electron neutrino has a
small mixing with the neutrinos whose mass-squared difference is
responsible for the oscillations of atmospheric neutrinos
($\nu_1$, $\nu_2$ in scheme A and $\nu_3$, $\nu_4$ in scheme B).
Therefore,
the transition probability of
electron neutrinos and antineutrinos
into other states
in atmospheric and long-baseline (LBL) experiments
is suppressed.
Indeed,
it can be shown~\cite{BGG97a}
that the transition probabilities
of electron neutrinos and antineutrinos
into all other states
and
the probability of
$
\stackrel{\makebox[0pt][l]
{$\hskip-3pt\scriptscriptstyle(-)$}}{\nu_{\mu}}
\to\stackrel{\makebox[0pt][l]
{$\hskip-3pt\scriptscriptstyle(-)$}}{\nu_{e}}
$
transitions in vacuum are bounded by
\begin{eqnarray}
&&
1 -
P^{({\rm LBL})}_{\stackrel{\makebox[0pt][l]
{$\hskip-3pt\scriptscriptstyle(-)$}}{\nu_{e}}
\to\stackrel{\makebox[0pt][l]
{$\hskip-3pt\scriptscriptstyle(-)$}}{\nu_{e}}}
\leq
a^{0}_{e}
\left( 2 - a^{0}_{e} \right)
\label{05}
\\
&&
P^{({\rm LBL})}_{\stackrel{\makebox[0pt][l]
{$\hskip-3pt\scriptscriptstyle(-)$}}{\nu_{\mu}}
\to\stackrel{\makebox[0pt][l]
{$\hskip-3pt\scriptscriptstyle(-)$}}{\nu_{e}}}
\leq
\min\!\left(
a^{0}_{e}
\left( 2 - a^{0}_{e} \right)
\, , \,
a^{0}_{e}
+
\frac{1}{4}
\,
A^{0}_{\mu;e}
\right)
\,
\label{06}
\end{eqnarray}
where
$A^{0}_{\mu;e}$
is the upper bound for the amplitude of
$
\stackrel{\makebox[0pt][l]
{$\hskip-3pt\scriptscriptstyle(-)$}}{\nu_{\mu}}
\to\stackrel{\makebox[0pt][l]
{$\hskip-3pt\scriptscriptstyle(-)$}}{\nu_{e}}
$
transitions measured in SBL experiments with accelerator neutrinos.

\begin{table}[t!]
\begin{tabular*}{\linewidth}{@{\extracolsep{\fill}}cc}
\begin{minipage}{0.47\linewidth}
\begin{center}
\mbox{\epsfig{file=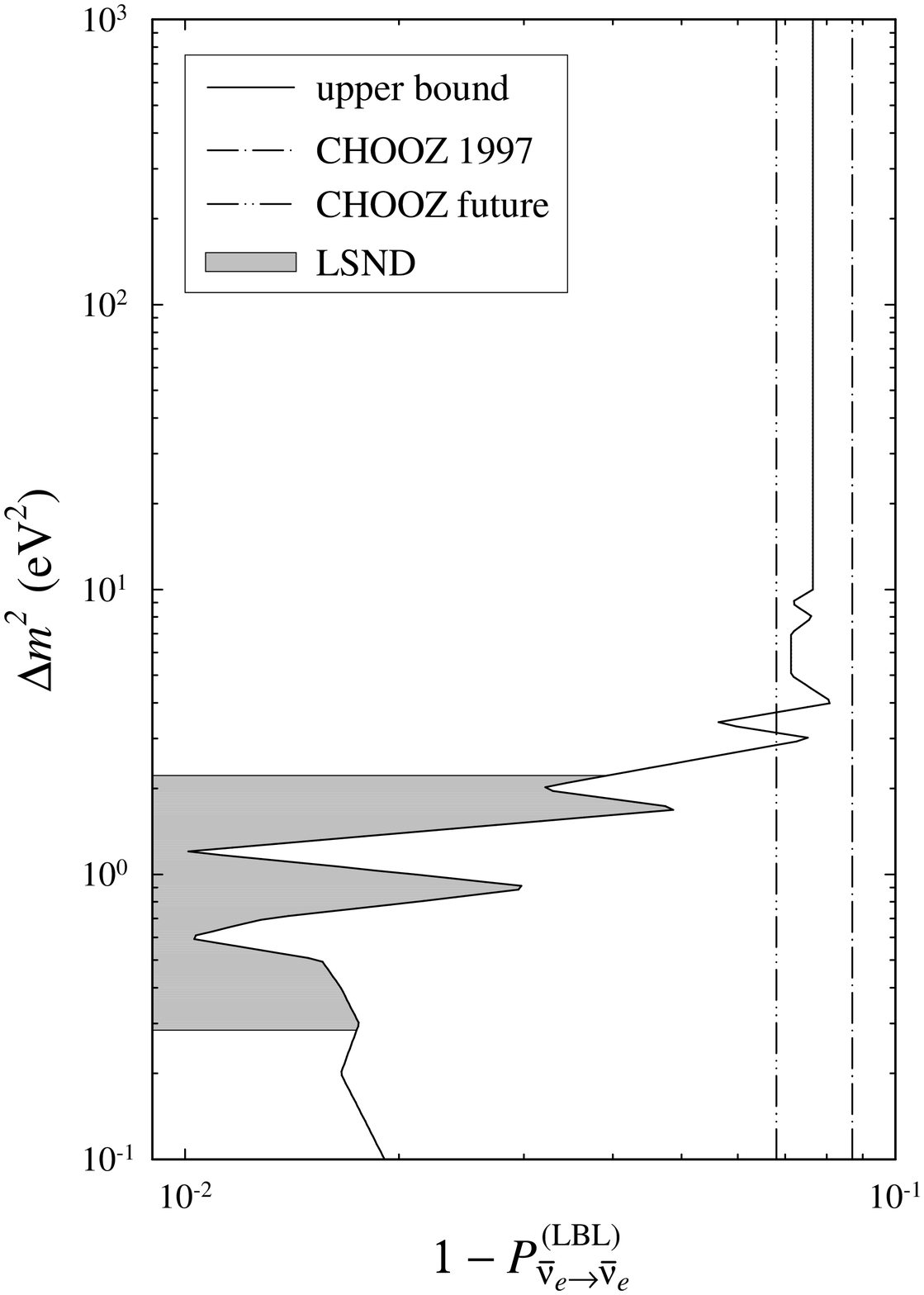,width=0.95\linewidth}}
\end{center}
\end{minipage}
&
\begin{minipage}{0.47\linewidth}
\begin{center}
\mbox{\epsfig{file=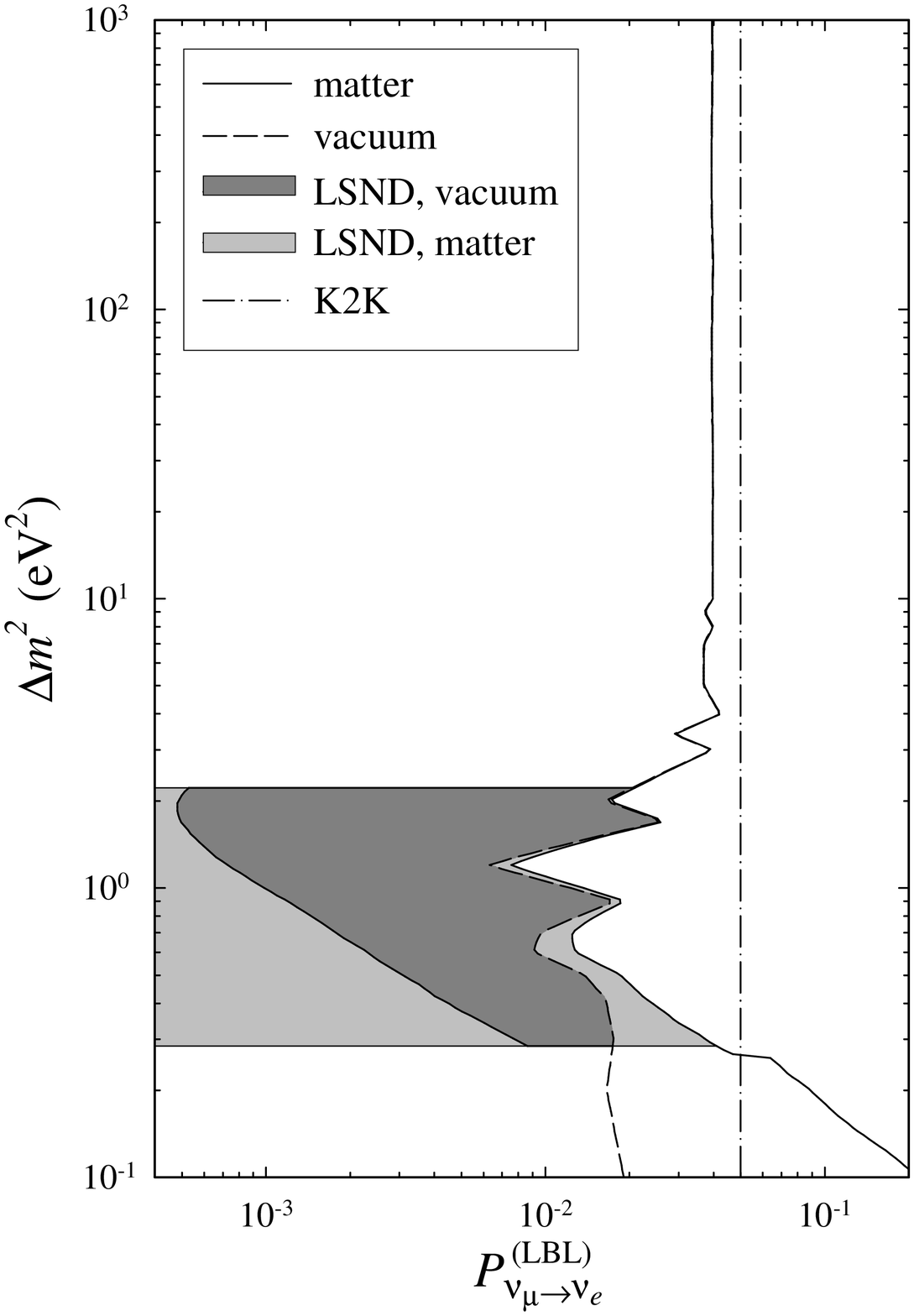,width=0.95\linewidth}}
\end{center}
\end{minipage}
\\
\refstepcounter{figure}
\label{fig1}                 
Figure \ref{fig1}
&
\refstepcounter{figure}
\label{fig2}                 
Figure \ref{fig2}
\end{tabular*}
\null \vspace{-0.5cm} \null
\end{table}

The upper bound (\ref{05}) obtained from the 90\% CL
exclusion plot of the Bugey
experiment~\cite{Bugey95} is shown
in Fig.~\ref{fig1}
(solid line).
The shadowed region in Fig.~\ref{fig1}
corresponds to the range of $\Delta{m}^2$
allowed at 90\% CL by the results of the LSND
and all the other SBL experiments.
The upper bound for the transition
probability of $\bar\nu_e$'s into other states
obtained from the recently published
exclusion plot of the CHOOZ~\cite{CHOOZ} experiment
is shown in Fig.~\ref{fig1} (dash-dotted line).
One can see that the CHOOZ upper bound
agrees with that obtained from Eq.(\ref{05}).
In Fig.~\ref{fig1} we have also drawn the curve corresponding
to the expected final sensitivity of the CHOOZ experiment
(dash-dot-dotted line).
One can see that
the LSND-allowed region imply an upper bound for
$1-P^{(\mathrm{LBL})}_{\bar \nu_e \to \bar \nu_e}$
of about
$ 5 \times 10^{-2} $,
which is smaller than the expected final sensitivity of
the CHOOZ experiment.

The bound (\ref{06}) obtained
from the 90\% CL
exclusion plots of the Bugey
experiment
and
of the
BNL E734, BNL E776 and CCFR
experiments~\cite{BNLE734-BNLE776-CCFR96}
is depicted by the dashed line
in Fig.~\ref{fig2}.
The solid line in Fig.~\ref{fig2} shows the upper bound on
$
P^{(\mathrm{LBL})}_{\stackrel{\makebox[0pt][l]
{$\hskip-3pt\scriptscriptstyle(-)$}}{\nu_{\mu}}
\to\stackrel{\makebox[0pt][l]
{$\hskip-3pt\scriptscriptstyle(-)$}}{\nu_{e}}}
$
taking into account matter effects.
The expected sensitivities
of the K2K long-baseline accelerator neutrino experiment~\cite{K2K}
is indicated in Fig.~\ref{fig2}
by the dash-dotted vertical line.
The shadowed region in Fig.~\ref{fig2}
corresponds to the range of $\Delta{m}^2$
allowed at 90\% CL by the results of the LSND
and all the other SBL experiments.
It can be seen that the results of SBL experiments
indicate an upper bound for
$
P^{(\mathrm{LBL})}_{\stackrel{\makebox[0pt][l]
{$\hskip-3pt\scriptscriptstyle(-)$}}{\nu_{\mu}}
\to\stackrel{\makebox[0pt][l]
{$\hskip-3pt\scriptscriptstyle(-)$}}{\nu_{e}}}
$
smaller than
$ 4 \times 10^{-2} $
and
smaller than the expected sensitivity of
the K2K experiment.

The two schemes A and B have
identical implications for neutrino oscillation experiments,
but very different implications
for neutrinoless double-$\beta$ decay experiments
and for tritium $\beta$-decay experiments.
Indeed,
in scheme A
\begin{equation}
|\langle{m}\rangle|
\leq
m_4
\qquad \mbox{and} \qquad
m(^3{\rm H}) \simeq m_4
\,,
\label{21}
\end{equation}
whereas in scheme B
\begin{equation}
|\langle{m}\rangle|
\leq
a_e^0 \, m_4
\ll
m_4
\qquad \mbox{and} \qquad
m(^3{\rm H}) \ll m_4
\,,
\label{22}
\end{equation}
where
$
\langle{m}\rangle
=
\sum_{i=1}^4 U_{ei}^2 \, m_i
$
is the effective Majorana mass
that determines the matrix element of neutrinoless double-$\beta$ decay
and
$m(^3{\rm H})$
is the neutrino mass measured in
tritium $\beta$-decay
experiments.
Therefore,
in scheme B
$|\langle{m}\rangle|$
and
$m(^3{\rm H})$
are smaller than the expected sensitivity
of the next generation of
neutrinoless double-$\beta$ decay
and tritium $\beta$-decay experiments.
The observation of a positive signal in these experiments
would be an indication in favor of scheme A.

\begin{figure}[t!]
\begin{center}
\setlength{\unitlength}{1cm}
\begin{picture}(13.8,4.4)
%
%
\put(0.0,4.4){\makebox(0,0)[lt]{\underline{Scheme A}}}
\put(0.0,2.0){\makebox(0,0)[l]{$\nu_\tau,\nu_s$?}}
\put(6.35,1.9){\makebox(1.1,0.0){$\Delta{m}^2_{{\rm SBL}}$}}
\put(13.8,2.0){\makebox(0,0)[r]{$\nu_\tau,\nu_s$?}}
\put(13.8,4.4){\makebox(0,0)[rt]{\underline{Scheme B}}}
\put(1.5,0.0){\begin{picture}(1.8,4.4)
\thicklines
\put(1.0,0.2){\vector(0,1){3.8}}
\put(1.0,4.1){\makebox(0,0)[lb]{$m$}}
\put(0.9,0.2){\line(1,0){0.2}}
\put(1.2,0.2){\makebox(0,0)[l]{$\nu_1$}}
\put(0.9,0.6){\line(1,0){0.2}}
\put(1.2,0.6){\makebox(0,0)[l]{$\nu_2$}}
\put(0.9,3.3){\line(1,0){0.2}}
\put(1.2,3.25){\makebox(0,0)[l]{$\nu_3$}}
\put(0.9,3.5){\line(1,0){0.2}}
\put(1.2,3.55){\makebox(0,0)[l]{$\nu_4$}}
\put(0.8,0.4){\makebox(0,0)[r]{$\nu_\mu$}}
\put(0.8,3.4){\makebox(0,0)[r]{$\nu_e$}}
\end{picture}}
\put(3.3,0.0){\begin{picture}(2.85,4.0)
\thinlines
\put(0.0,0.6){\line(2,-1){0.4}}
\put(0.0,0.2){\line(2, 1){0.4}}
\put(0.6,0.4){\makebox(0,0)[l]{$\Delta{m}^2_{{\rm atm}}$}}
\put(0.0,3.6){\line(2,-1){0.4}}
\put(0.0,3.2){\line(2, 1){0.4}}
\put(0.6,3.4){\makebox(0,0)[l]{$\Delta{m}^2_{{\rm sun}}$}}
\put(2.0,3.6){\line(1,-2){0.85}}
\put(2.0,0.2){\line(1, 2){0.85}}
\end{picture}}
\put(7.65,0.0){\begin{picture}(2.85,4.0)
\thinlines
\put(2.85,0.5){\line(-2,-1){0.4}}
\put(2.85,0.1){\line(-2, 1){0.4}}
\put(2.25,0.3){\makebox(0,0)[r]{$\Delta{m}^2_{{\rm sun}}$}}
\put(2.85,3.6){\line(-2,-1){0.4}}
\put(2.85,3.2){\line(-2, 1){0.4}}
\put(2.25,3.4){\makebox(0,0)[r]{$\Delta{m}^2_{{\rm atm}}$}}
\put(0.85,3.6){\line(-1,-2){0.85}}
\put(0.85,0.2){\line(-1, 2){0.85}}
\end{picture}}
\put(10.50,0.0){\begin{picture}(1.8,4.4)
\thicklines
\put(0.8,0.2){\vector(0,1){3.8}}
\put(0.8,4.1){\makebox(0,0)[rb]{$m$}}
\put(0.7,0.2){\line(1,0){0.2}}
\put(0.6,0.15){\makebox(0,0)[r]{$\nu_1$}}
\put(0.7,0.4){\line(1,0){0.2}}
\put(0.6,0.45){\makebox(0,0)[r]{$\nu_2$}}
\put(0.7,3.1){\line(1,0){0.2}}
\put(0.6,3.1){\makebox(0,0)[r]{$\nu_3$}}
\put(0.7,3.5){\line(1,0){0.2}}
\put(0.6,3.5){\makebox(0,0)[r]{$\nu_4$}}
\put(1.0,0.3){\makebox(0,0)[l]{$\nu_e$}}
\put(1.0,3.3){\makebox(0,0)[l]{$\nu_\mu$}}
\end{picture}}
\end{picture}
\end{center}
\begin{center}
Figure \ref{fig3}
\end{center}
\null \vspace{-1.5cm} \null
\refstepcounter{figure}
\label{fig3}
\end{figure}

Summarizing,
the results of neutrino oscillation experiments
indicate that only the two four-neutrino schemes
(\ref{AB}) are allowed
and the electron neutrino has a
very small mixing
with the two massive neutrinos
that are responsible for the oscillations of atmospheric neutrinos
($\nu_1$, $\nu_2$ in scheme A
and
$\nu_3$, $\nu_4$ in scheme B).
Hence,
the two schemes have the form shown in Fig.~\ref{fig3},
where $\nu_e$ is associated with
the two massive neutrinos neutrinos
that are responsible for the oscillations of solar neutrinos
($\nu_3$, $\nu_4$ in scheme A
and
$\nu_1$, $\nu_2$ in scheme B),
with which it has a large mixing,
whereas
$\nu_\mu$ is associated with
the two massive neutrinos neutrinos
that are responsible for the oscillations of atmospheric neutrinos,
with which the muon neutrino has a large mixing.
The results of neutrino oscillation experiments
do not provide yet an indication of
where $\nu_\tau$ and $\nu_s$ have to be placed
in the two schemes represented in Fig.~\ref{fig3}.
In the following Section we will show that
the standard Big-Bang Nucleosynthesis constraint
on the number of light neutrinos
provide a stringent limit on the mixing of the sterile neutrino
with the two massive neutrinos
that are responsible for the oscillations
of atmospheric neutrinos~\cite{OY96,BGGS98}.

\section{BBN constraints on 4-neutrino mixing}
\label{BBN constraints on 4-neutrino mixing}

It is well known that
the observed abundance of primordial light elements
is predicted with an impressive degree of accuracy by
the standard model of Big-Bang Nucleosynthesis
if the number $N_\nu$
of light neutrinos (with mass much smaller than 1 MeV)
in equilibrium at the neutrino decoupling temperature
($ T_{{\rm dec}} \simeq 2 \, {\rm MeV} $
for $\nu_e$
and
$ T_{{\rm dec}} \simeq 4 \, {\rm MeV} $
for $\nu_\mu$, $\nu_\tau$)
is not far from three~\cite{ST98,Olive97}.

The value of $N_\nu$
is especially crucial for the primordial abundance of $^4$He.
This is due to the fact that $N_\nu$
determines the freeze-out temperature
of the weak interaction processes
$ e^+ + n \leftrightarrows p + \bar\nu_e $,
$ \nu_e + n \leftrightarrows p + e^- $
and
$ n \leftrightarrows p + e^- + \bar\nu_e $
that maintain protons and neutrons in equilibrium,
\textit{i.e.}
the temperature at which the rate
$
\Gamma_{W}(T) \simeq G_{F} T^5
$
($G_F$ is the Fermi constant)
of these weak interaction processes
becomes smaller than the expansion rate of the universe
\begin{equation}
H(T) \equiv \frac{\dot{R}(T)}{R(T)}
= \sqrt{ \frac{ 8 \pi^3 }{ 90 } \, g_* } \, \frac{ T^2 }{ M_P }
\label{904}
\end{equation}
($M_P$ is the Planck mass),
where $R(T)$ is the cosmic scale factor
and
$ g_* = 5.5 + 1.75 N_\nu $ for $ m_e \lesssim T \lesssim m_\mu $.
If
$N_\nu=3$
the primordial mass fraction of $^4$He,
$ Y_P \equiv \mbox{mass density of $^4$He / total mass density} $,
is
$ Y_P \simeq 0.24 $,
which agrees very well with the observed value~\cite{Olive97}
$ Y_P^{{\rm obs}} = 0.238 \pm 0.002 $.
Since
$Y_P$
is very sensitive to variations of $N_\nu$,
it is clear that the observed value of $Y_P$
puts stringent constraints on the possible deviation of
$N_\nu$ from the Standard Model value
$N_\nu=3$.

In the four-neutrino schemes (\ref{AB})
standard BBN gives a constraint on neutrino mixing if the upper bound for
$N_\nu$ is less than four.
In this case the mixing of the sterile neutrino is severely constrained
because otherwise neutrino oscillations
would bring the sterile neutrino in equilibrium
before neutrino decoupling
\cite{kainulainen,barbieri,resonance,enqvist91,enqvist92,cline,shi,OY96},
leading to $N_\nu=4$.
In particular,
we will show that standard BBN with $N_\nu<4$
implies that $c_s$ is extremely small.

The amount of sterile neutrinos present at nucleosynthesis can be
calculated using the differential equation~\cite{kainulainen}
\begin{equation}
\frac{{\rm d}n_{\nu_s}}{{\rm d}T}
=
- \frac{1}{2HT}
\sum_{\alpha=e,\mu,\tau}
\langle P_{\nu_\alpha\to\nu_s} \rangle_{{\rm coll}}
\Gamma_{\nu_\alpha}
\ 
(1-n_{\nu_s})
\,,
\label{master}
\end{equation}
where $n_{\nu_s}$ is the number density of the sterile neutrino relative
to the number density of an active neutrino in equilibrium and
$\Gamma_{\nu_\alpha}$ are the collision rates of the active neutrinos,
including elastic and inelastic scattering~\cite{enqvist92},
$\Gamma_{\nu_e} = 4.0 \, G_F^2 T^5$
and
$\Gamma_{\nu_\mu} = \Gamma_{\nu_\tau} = 0.7 \, \Gamma_{\nu_e}$.
The quantities
$ \langle P_{\nu_\alpha\to\nu_s} \rangle_{{\rm coll}} $
are the probabilities of $\nu_\alpha\to\nu_s$ transitions
averaged over the collision time
$t_{{\rm coll}} = 1/\Gamma_{\nu_\alpha}$.
Hence,
also $n_{\nu_s}$ has to be considered as a quantity
averaged over the collision time.

Equation (\ref{master})
describes non-resonant and adiabatic resonant
neutrino transitions
if
$t_{{\rm osc}} \ll t_{{\rm coll}} \ll t_{{\rm exp}}$.
The
condition $t_{{\rm osc}} \ll t_{{\rm coll}}$ means that neutrino
oscillations have to be fast relative to the
collision time.
The characteristic expansion time of the universe
$t_{{\rm exp}}$ is given by
$t_{{\rm exp}} = 1/H$
where $H$ is the Hubble parameter
$ H \equiv \dot{R}/R $,
which is related to the temperature $T$ by
$H=-\dot{T}/T \simeq 0.7 \, (T/1\,{\rm MeV})^2 \,{\rm s}^{-1}$
(this value can be obtained from Eq.(\ref{904}) with
$ m_e \lesssim T \lesssim m_\mu $ and $N_\nu\simeq3$).
The relation
$\Gamma_{\nu_e}/H \simeq 1.2\, (T/1\, {\rm MeV})^3$ shows that for
temperatures larger than 2 MeV the collision time is always much
smaller than the expansion time~\cite{kainulainen}.

Since by definition $N_\nu$
is the effective number of massless neutrino species
at $T_{{\rm dec}}$,
in order to get a constraint on the mixing of sterile neutrinos
we need to calculate
the value of $n_{\nu_s}$ at $T_{{\rm dec}}$
produced by neutrino oscillations.
With the initial condition
$n_{\nu_s}(T_i)=0$ ($T_i\sim 100$ MeV),
the integration of
Eq.(\ref{master}) gives~\cite{cline}
$ n_{\nu_s}(T_{{\rm dec}}) = 1 - e^{-F} $
with
\begin{equation}
F
=
\int_{T_{{\rm dec}}}^{T_i} \frac{1}{2HT}
\sum_{\alpha=e,\mu,\tau}
\langle P_{\nu_\alpha\to\nu_s} \rangle_{{\rm coll}}
\Gamma_{\nu_\alpha}
{\rm d}T
\,.
\label{F}
\end{equation}
Imposing the upper bound
$ n_{\nu_s}(T_{{\rm dec}}) \leq \delta N \equiv N_\nu - 3 $
one obtains the condition
$ F \leq |\ln(1-\delta N)| $.

\begin{table}[t!]
\begin{tabular*}{\linewidth}{@{\extracolsep{\fill}}cc}
\begin{minipage}{0.47\linewidth}
\begin{center}
\mbox{\epsfig{file=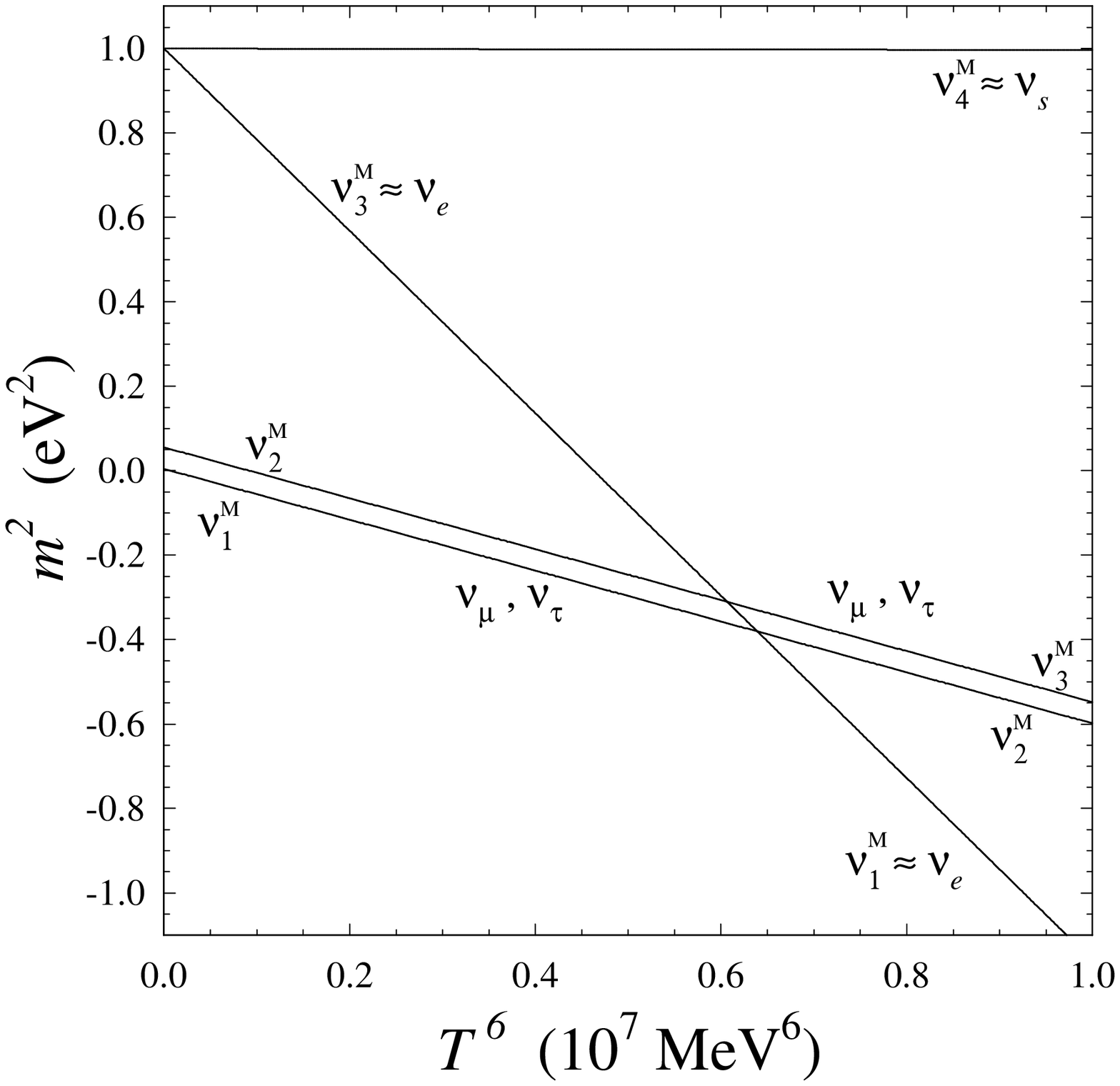,width=0.95\linewidth}}
\end{center}
\end{minipage}
&
\begin{minipage}{0.47\linewidth}
\begin{center}
\mbox{\epsfig{file=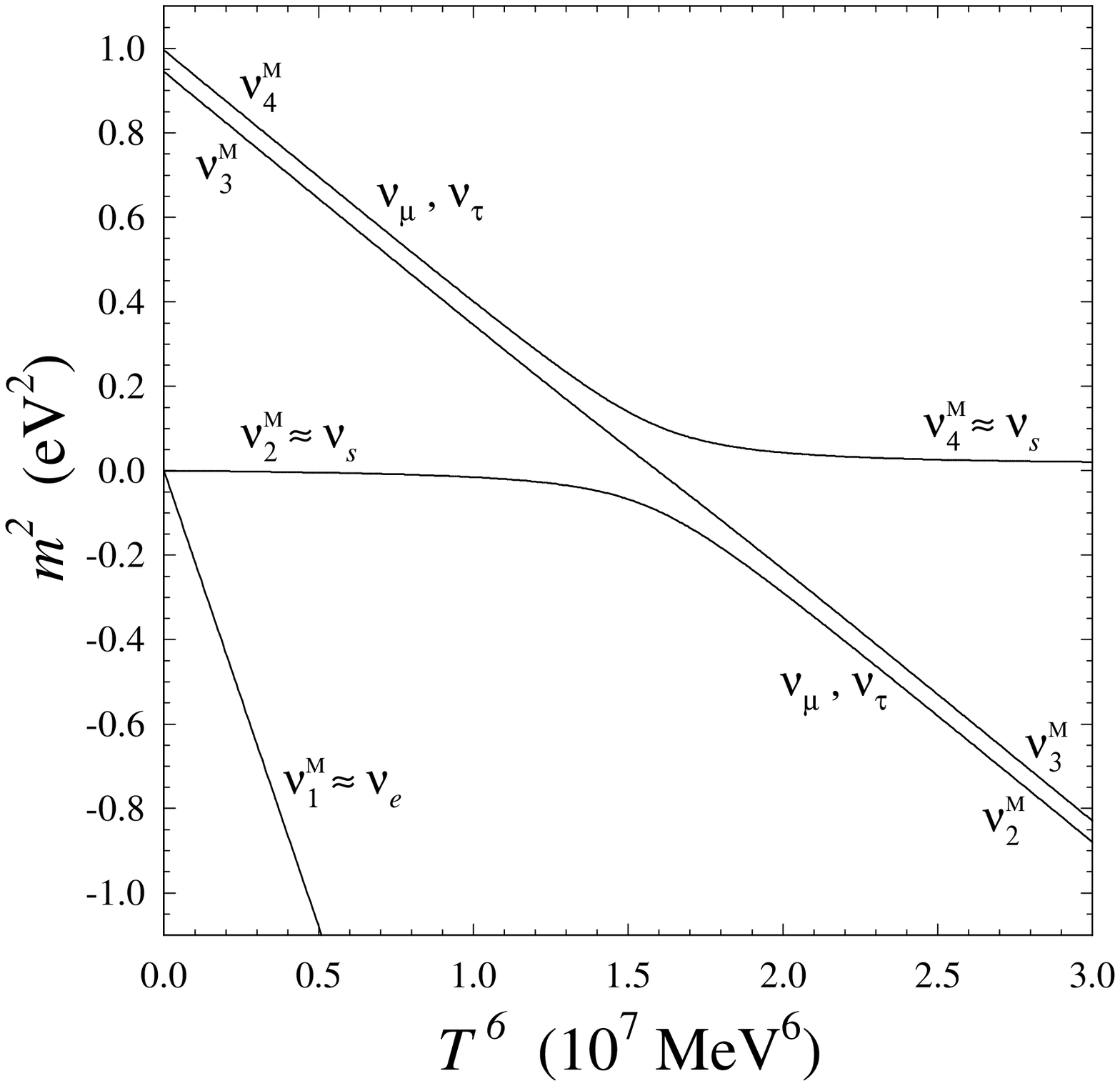,width=0.95\linewidth}}
\end{center}
\end{minipage}
\\
\refstepcounter{figure}
\label{fig4}                 
Figure \ref{fig4}
&
\refstepcounter{figure}
\label{fig5}                 
Figure \ref{fig5}
\end{tabular*}
\null \vspace{-0.5cm} \null
\end{table}

For the calculation of $F$
the averaged transition probabilities
$ \langle P_{\nu_\alpha\to\nu_s} \rangle_{{\rm coll}} $
must be evaluated
and the effective potentials of neutrinos and antineutrinos
due to coherent forward scattering
in the primordial plasma~\cite{raffelt},
\begin{equation}
V_e = - 6.02 \, G_F \, p \, \frac{T^4}{M_W^2} \equiv V
\,,
\quad
V_{\mu,\tau} = \xi V
\quad \mbox{and} \quad
V_s=0
\,,
\end{equation}
must be taken into account
(in the absence of a
lepton asymmetry
the effective potentials of neutrinos and antineutrinos
are equal).
Here $p$ is the neutrino momentum,
which we approximate with
its temperature average $ \langle p \rangle \simeq 3.15 \, T $,
$G_F$ is the Fermi constant,
$M_W$ is the mass of the $W$ boson
and
$\xi = \cos^2 \theta_W / (2+\cos^2 \theta_W) \simeq 0.28$,
where $\theta_W$ is the weak mixing angle.
The propagation of neutrinos and antineutrinos
is governed by the effective hamiltonian
in the weak basis
\begin{equation}
H_W
=
p
+
\frac{1}{2p}
\, U \, {\rm diag}\bigg[m_1^2,m_2^2,m_3^2,m_4^2\bigg] \, U^{\dagger}
+ {\rm diag}\bigg[V,\xi V,\xi V,0\bigg]
\,.
\label{HW}
\end{equation}
It is convenient to subtract from $H_W$
the constant term
$ p + m_1/2p + \xi V $,
which does not affect the relative evolution of the neutrino flavor states,
in order to get
\begin{equation}
H'_W
=
\frac{1}{2p} \, U \,
{\rm diag}\bigg[0,\Delta{m}^2_{21},\Delta{m}^2_{31},\Delta{m}^2_{41}\bigg]
\, U^{\dagger}
+ {\rm diag}\bigg[(1-\xi)V,0,0,-\xi V\bigg]
\,.
\label{HPW}
\end{equation}
From this expression it is clear that the relative evolution
of the flavor neutrino states
depends on the three mass-squared differences
and not on the absolute scale of the neutrino masses.
The effective hamiltonian in the mass basis is given by
$ H'_M = U^{\dagger} \, H'_W \, U $:
\begin{equation}
H'_M
=
\frac{1}{2p} \, {\rm diag}\bigg[0,\Delta{m}^2_{21},\Delta{m}^2_{31},\Delta{m}^2_{41}\bigg]
+ U^{\dagger} \, {\rm diag}\bigg[(1-\xi)V,0,0,-\xi V\bigg] \, U
\,.
\label{HPM}
\end{equation}
In the mass basis
the mixing has been transferred from the mass term to the potential term.
In order to calculate the evolution of the neutrino flavors
it is necessary to parameterize the $4\times4$
neutrino mixing matrix $U$.
However,
since
the second and third rows
and
the second and third columns
of the diagonal potential matrix in Eq.(\ref{HPM})
are equal to zero, it is clear that
the values of the second and third rows of $U$,
corresponding to $\nu_\mu$ and $\nu_\tau$,
are irrelevant and do not need to be parameterized.
Furthermore,
since $c_e$ is small in both schemes A and B,
it does not have any effect on neutrino oscillations
before BBN
and the approximation
$c_e=0$
is allowed.
Hence,
the $4\times4$ neutrino mixing matrix
can be partially parameterized as
\begin{equation}
U
=
\left(
\begin{array}{cccc} \displaystyle
0 & 0 & \cos\theta & \sin\theta
\\ \displaystyle
\cdot & \cdot & \cdot & \cdot
\\ \displaystyle
\cdot & \cdot & \cdot & \cdot
\\ \displaystyle
\sin\varphi \sin\chi & -\sin\varphi \cos\chi &
-\cos\varphi \sin\theta & \cos\varphi \cos\theta
\end{array}
\right)
\label{PA}
\end{equation}
in scheme A
and
\begin{equation}
U
=
\left(
\begin{array}{cccc} \displaystyle
\cos\theta & \sin\theta & 0 & 0
\\ \displaystyle
\cdot & \cdot & \cdot & \cdot
\\ \displaystyle
\cdot & \cdot & \cdot & \cdot
\\ \displaystyle
- \cos\varphi \sin\theta & \cos\varphi \cos\theta &
\sin\varphi \sin\chi & -\sin\varphi \cos\chi
\end{array}
\right)
\label{PB}
\end{equation}
in scheme B,
with $0 \leq \varphi \leq \pi/2$.
Notice that
$ c_s = \sin^2 \varphi $
in both schemes A and B.
The dots in Eqs.(\ref{PA}) and (\ref{PB})
indicate the elements of the mixing matrix
belonging to the $\nu_\mu$ and $\nu_\tau$ rows
($U_{{\mu}i}$ and $U_{{\tau}i}$ with $i=1,\ldots,4$),
which do not need to be parameterized.
In Eqs.(\ref{PA}) and (\ref{PB})
we have parameterized only the elements
of the mixing matrix
belonging to the $\nu_e$ and $\nu_s$ lines
($U_{ei}$ and $U_{si}$ with $i=1,\ldots,4$)
in terms of the three mixing angles
$\theta$, $\chi$, $\varphi$.
It is clear that this partial parameterization
of the mixing matrix
(with the approximation
$U_{e1}=U_{e2}=0$)
is much easier to manipulate than a complete
parameterization,
which would require the introduction of
6 mixing angles and 3 complex phases.

\begin{table}[t!]
\begin{tabular*}{\linewidth}{@{\extracolsep{\fill}}cc}
\begin{minipage}{0.47\linewidth}
\begin{center}
\mbox{\epsfig{file=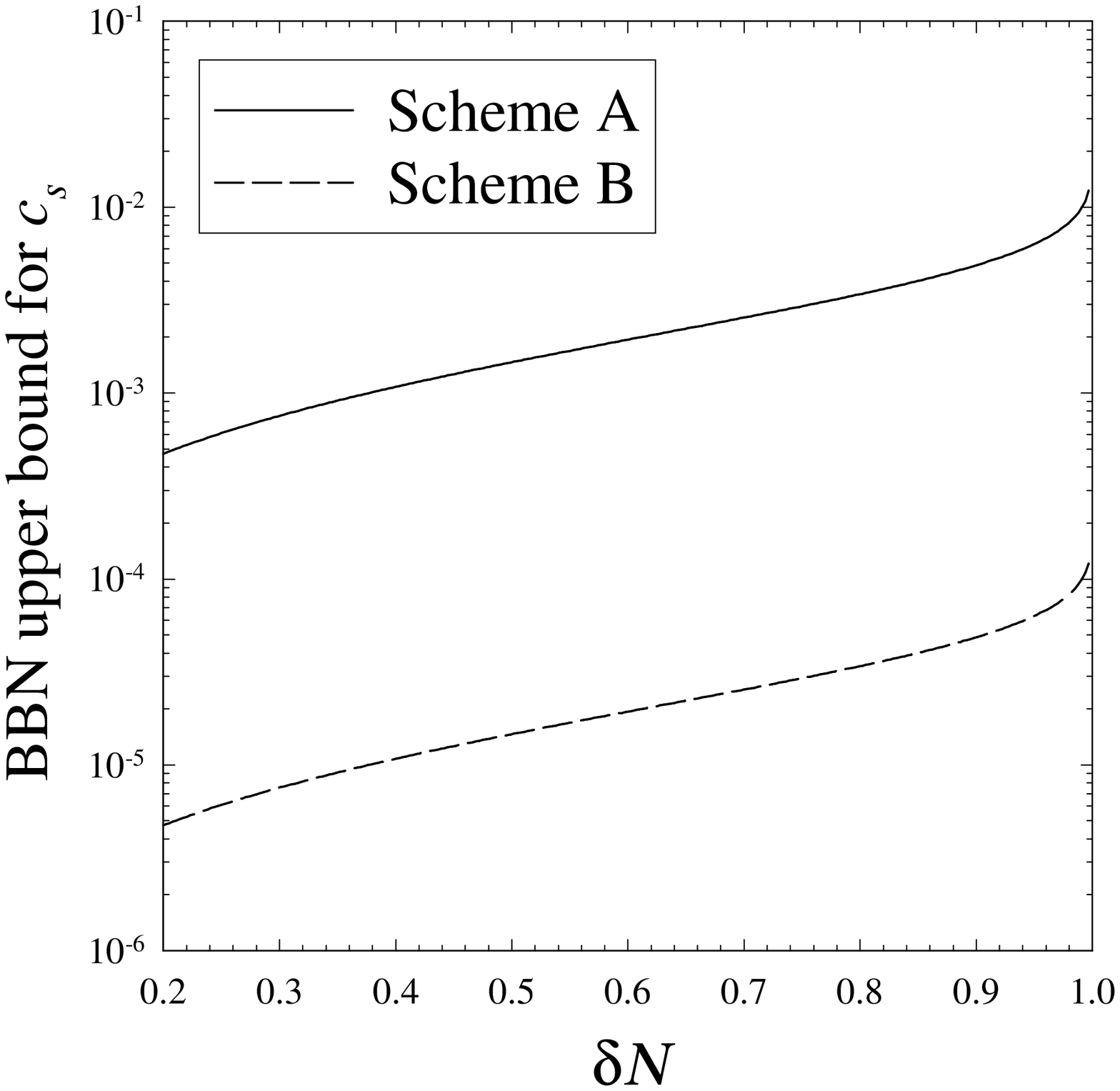,width=0.95\linewidth}}
\end{center}
\end{minipage}
&
\begin{minipage}{0.47\linewidth}
\begin{center}
\mbox{\epsfig{file=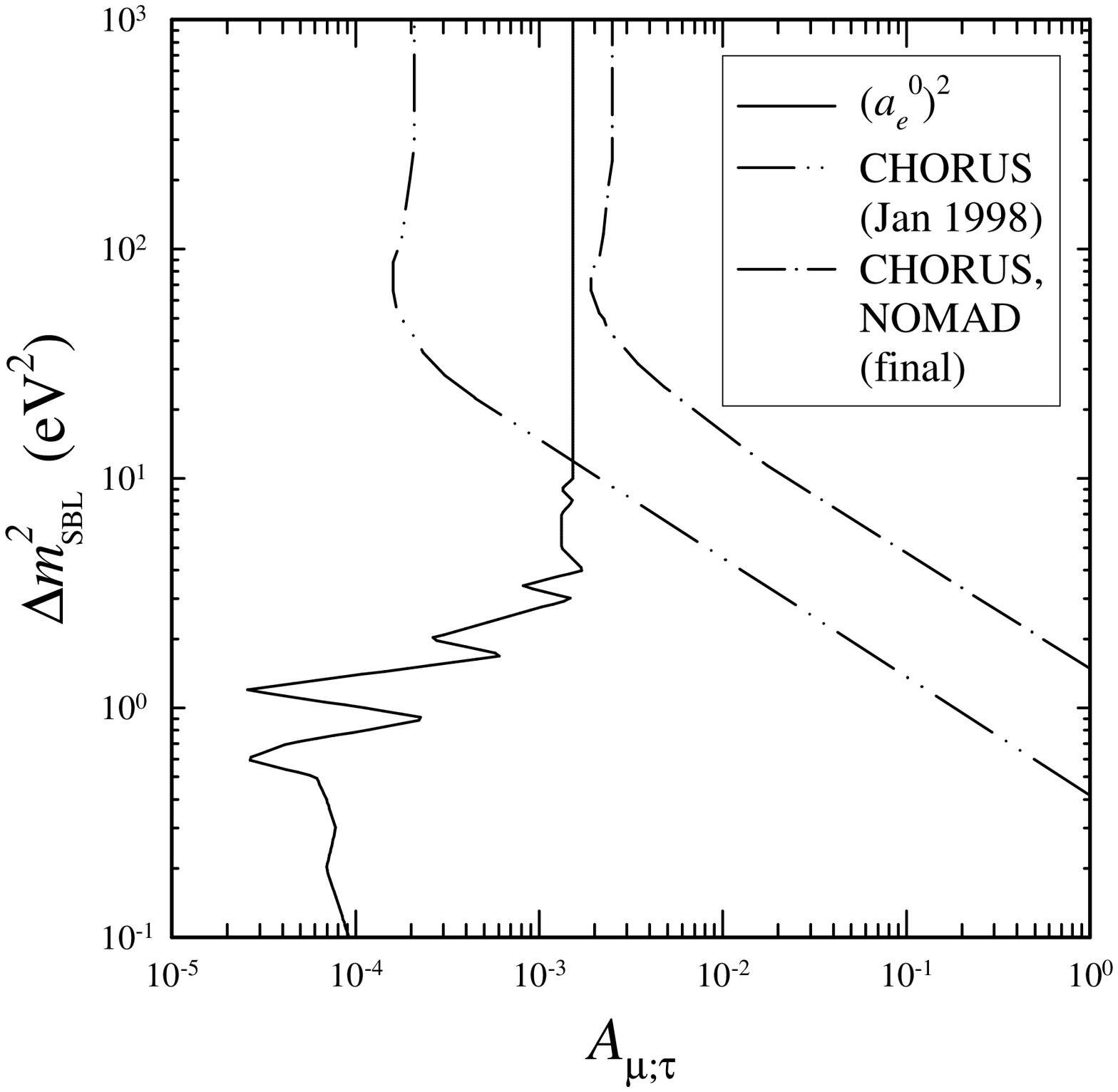,width=0.95\linewidth}}
\end{center}
\end{minipage}
\\
\refstepcounter{figure}
\label{fig6}                 
Figure \ref{fig6}
&
\refstepcounter{figure}
\label{fig7}                 
Figure \ref{fig7}
\end{tabular*}
\null \vspace{-0.5cm} \null
\end{table}

Notice that no complex phase is needed for the
partial parameterization of the mixing matrix in Eqs.(\ref{PA}) and (\ref{PB}),
because the elements
$U_{ei}$ and $U_{si}$ with $i=1,\ldots,4$
can be chosen real.
Indeed,
if we consider,
for example,
the scheme A,
the line
$U_{si}$ with $i=1,\ldots,4$
and the element $U_{e4}$
can be chosen real
because all
observable transition probabilities are invariant under the
phase transformation
$
U_{{\alpha}j}
\to
e^{i x_\alpha}
\,
U_{{\alpha}j}
\,
e^{i y_j}
$,
where $x_\alpha$ and $y_j$ are arbitrary parameters.
In order to show that also $U_{e3}$
can be chosen real,
we multiply the unitarity relation
$ \sum_{k=1}^{4} U_{ek} \, U_{sk}^* = 0 $
with
$ U_{e3}^* \, U_{s3} $.
The imaginary part of the resulting relation gives
\begin{equation}
{\rm Im}\big[ U_{e3} \, U_{s3}^* \, U_{e4}^* \, U_{s4} \big]
=
{\rm Im}\big[ U_{e1} \, U_{s1}^* \, U_{e3}^* \, U_{s3} \big]
+
{\rm Im}\big[ U_{e2} \, U_{s2}^* \, U_{e3}^* \, U_{s3} \big]
\,.
\label{452}
\end{equation}
In the approximation $ U_{e1} = U_{e2} = 0 $
the right-hand part of Eq.(\ref{452}) vanishes.
Therefore,
since
$U_{s3}$, $U_{e4}$, $U_{s4}$
have been chosen to be real,
also $U_{e3}$ must be real.

Since the mass-squared differences have a hierarchical structure,
$\Delta{m}^2_{43}\ll\Delta{m}^2_{21}\ll\Delta{m}^2_{41}$
in scheme A
and
$\Delta{m}^2_{21}\ll\Delta{m}^2_{43}\ll\Delta{m}^2_{41}$
in scheme B,
the effective hamiltonian $H'_M$
can be diagonalized approximately
taking into account only one of the three
$\Delta{m}^2$'s
for different ranges of the temperature $T$.
Then, it can be shown that~\cite{OY96,BGGS98}
the condition
$ F \leq |\ln(1-\delta N)| $
gives the bound
\begin{equation}\label{bound}
920 \left( \frac{\Delta{m}^2_{{\rm SBL}}}{1\, {\rm eV}^2} \right)^{1/2}
d_s \, \sqrt{1-d_s}
+
33 \left( \frac{\Delta{m}^2_{{\rm atm}}}{10^{-2} \,
{\rm eV}^2} \right)^{1/2}
\frac{\sin^2 2\chi}{\sqrt{1+\cos 2\chi}} \, c_s^{3/2}
\leq |\ln(1-\delta N)|
\,,
\end{equation}
with
$ d_s \equiv c_s $
in scheme A
and
$ d_s \equiv 1 - c_s $
in scheme B.

Both terms in the left-hand side of Eq.(\ref{bound})
are positive and must be small
if $ \delta N < 1 $.
The SBL term,
depending on
$\Delta{m}^2_{{\rm SBL}}$,
is small if either $c_s$ is small or large,
but the atmospheric term,
which depends on
$\Delta{m}^2_{{\rm atm}}$,
is small only if $c_s$ is small.
Indeed,
if $c_s$ is close to one we have
$(U_{\mu 1},U_{\mu 2}) \sim (\cos \chi, \sin \chi)$
in scheme A
and
$(U_{\mu 3},U_{\mu 4}) \sim (\cos \chi, \sin \chi)$
in scheme B.
This means that,
in order to accommodate
the atmospheric neutrino anomaly,
$\sin^2 2\chi$ cannot be small.
This is in contradiction
with the inequality (\ref{bound}) and
we conclude that the bound (\ref{bound})
implies that $c_s$ is small.

Since $c_s$ is small
only non-resonant transitions
of active into sterile neutrinos due to
$\Delta{m}^2_{{\rm SBL}}$
are possible in scheme A,
as illustrated in Fig.~\ref{fig4}
where we have plotted
the effective squared masses
(obtained from a numerical diagonalization of the hamiltonian (\ref{HW}))
as functions of $T^6$
($\nu_e$ does not have resonant transitions into $\nu_\mu$ or $\nu_\tau$
because we have chosen $c_e=0$).
Hence,
the conditions for the validity of Eq.(\ref{master}) are satisfied
and
the SBL term in Eq.(\ref{bound}) gives the bound
\begin{equation}
c_s \leq 1.1 \times 10^{-3}
\left( \frac{\Delta{m}^2_{{\rm SBL}}}{1\, {\rm eV}^2} \right)^{-1/2}
|\ln(1-\delta N)|
\,.
\label{boundA}
\end{equation}
On the other hand,
since $c_s$ is small,
a resonance occurs in scheme B at the temperature
$
T_{{\rm res}}
=
16 (\Delta{m}^2_{{\rm SBL}} / 1\,{\rm eV}^2)^{1/6} |1-2c_s|^{1/6} \, {\rm MeV}
$,
as illustrated in Fig.~\ref{fig5}.
The condition
$ \delta N < 1 $
implies that this resonance must not be passed adiabatically.
In this case
the conditions for the validity of Eq.(\ref{master}) are not fulfilled
and
the SBL term of Eq.(\ref{bound}) does not apply.
Using an appropriate formula~\cite{resonance}
for the calculation of the amount of sterile neutrinos produced at
the resonance through non-adiabatic transitions
one can show~\cite{BGGS98}
that the BBN bound on $c_s$ in scheme B is given by
\begin{equation}
c_s \leq 1.1 \times 10^{-5}
\left( \frac{\Delta{m}^2_{{\rm SBL}}}{1\:{\rm eV}^2}\right)^{-1/2}
|\ln(1-\delta N)|
\,.
\label{boundB}
\end{equation}

\begin{figure}[t!]
\begin{center}
\setlength{\unitlength}{1cm}
\begin{picture}(13.8,4.4)
%
%
\put(0.0,4.4){\makebox(0,0)[lt]{\underline{Scheme A}}}
\put(6.35,1.9){\makebox(1.1,0.0){$\Delta{m}^2_{{\rm SBL}}$}}
\put(13.8,4.4){\makebox(0,0)[rt]{\underline{Scheme B}}}
\put(1.5,0.0){\begin{picture}(1.8,4.4)
\thicklines
\put(1.0,0.2){\vector(0,1){3.8}}
\put(1.0,4.1){\makebox(0,0)[lb]{$m$}}
\put(0.9,0.2){\line(1,0){0.2}}
\put(1.2,0.2){\makebox(0,0)[l]{$\nu_1$}}
\put(0.9,0.6){\line(1,0){0.2}}
\put(1.2,0.6){\makebox(0,0)[l]{$\nu_2$}}
\put(0.9,3.3){\line(1,0){0.2}}
\put(1.2,3.25){\makebox(0,0)[l]{$\nu_3$}}
\put(0.9,3.5){\line(1,0){0.2}}
\put(1.2,3.55){\makebox(0,0)[l]{$\nu_4$}}
\put(0.8,0.4){\makebox(0,0)[r]{$\nu_\mu,\nu_\tau$}}
\put(0.8,3.4){\makebox(0,0)[r]{$\nu_e,\nu_s$}}
\end{picture}}
\put(3.3,0.0){\begin{picture}(2.85,4.0)
\thinlines
\put(0.0,0.6){\line(2,-1){0.4}}
\put(0.0,0.2){\line(2, 1){0.4}}
\put(0.6,0.4){\makebox(0,0)[l]{$\Delta{m}^2_{{\rm atm}}$}}
\put(0.0,3.6){\line(2,-1){0.4}}
\put(0.0,3.2){\line(2, 1){0.4}}
\put(0.6,3.4){\makebox(0,0)[l]{$\Delta{m}^2_{{\rm sun}}$}}
\put(2.0,3.6){\line(1,-2){0.85}}
\put(2.0,0.2){\line(1, 2){0.85}}
\end{picture}}
\put(7.65,0.0){\begin{picture}(2.85,4.0)
\thinlines
\put(2.85,0.5){\line(-2,-1){0.4}}
\put(2.85,0.1){\line(-2, 1){0.4}}
\put(2.25,0.3){\makebox(0,0)[r]{$\Delta{m}^2_{{\rm sun}}$}}
\put(2.85,3.6){\line(-2,-1){0.4}}
\put(2.85,3.2){\line(-2, 1){0.4}}
\put(2.25,3.4){\makebox(0,0)[r]{$\Delta{m}^2_{{\rm atm}}$}}
\put(0.85,3.6){\line(-1,-2){0.85}}
\put(0.85,0.2){\line(-1, 2){0.85}}
\end{picture}}
\put(10.50,0.0){\begin{picture}(1.8,4.4)
\thicklines
\put(0.8,0.2){\vector(0,1){3.8}}
\put(0.8,4.1){\makebox(0,0)[rb]{$m$}}
\put(0.7,0.2){\line(1,0){0.2}}
\put(0.6,0.15){\makebox(0,0)[r]{$\nu_1$}}
\put(0.7,0.4){\line(1,0){0.2}}
\put(0.6,0.45){\makebox(0,0)[r]{$\nu_2$}}
\put(0.7,3.1){\line(1,0){0.2}}
\put(0.6,3.1){\makebox(0,0)[r]{$\nu_3$}}
\put(0.7,3.5){\line(1,0){0.2}}
\put(0.6,3.5){\makebox(0,0)[r]{$\nu_4$}}
\put(1.0,0.3){\makebox(0,0)[l]{$\nu_e,\nu_s$}}
\put(1.0,3.3){\makebox(0,0)[l]{$\nu_\mu,\nu_\tau$}}
\end{picture}}
\end{picture}
\end{center}
\begin{center}
Figure \ref{fig8}
\end{center}
\null \vspace{-1.5cm} \null
\refstepcounter{figure}
\label{fig8}
\end{figure}

Figure \ref{fig6}
shows the values of the bounds
(\ref{boundA}) and (\ref{boundB})
obtained from the LSND~\cite{LSND} lower bound
$\Delta{m}^2_{{\rm SBL}} \gtrsim 0.27 \, {\rm eV}^2$
for $0.2\leq\delta N<1$.
One can see that
standard BBN implies that
$c_s$ is extremely small.

\section{Conclusions}
\label{Conclusions}

The BBN constraint on $c_s$ derived in the previous Section
implies that
$\nu_s$ is mainly mixed with the
two massive neutrinos that contribute to solar neutrino oscillations
($\nu_3$ and $\nu_4$ in scheme A
and
$\nu_1$ and $\nu_2$ in scheme B)
and the unitarity of the mixing matrix implies that
$\nu_\tau$ is mainly mixed with the
two massive neutrinos that contribute to the oscillations
of atmospheric neutrinos.
Adding this information to the two schemes of neutrino mixing
depicted in Fig.~\ref{fig3}
we obtain the schemes shown in Fig.~\ref{fig8}.

The two neutrino mixing schemes shown in Fig.~\ref{fig8}
have the following testable implications
for solar, atmospheric, long-baseline and
short-baseline neutrino oscillation experiments:
\begin{itemize}
\item
The solar neutrino problem
is due to
$\nu_e\to\nu_s$
oscillations.
This prediction will be checked by future solar neutrino
experiments
that can measure the ratio of
neutral-current and charged-current events~\cite{BG95}.
\item
The atmospheric neutrino anomaly is due to
$\nu_\mu\to\nu_\tau$
oscillations.
This prediction will be checked by LBL experiments.
Furthermore,
the absence of
$\nu_\mu\to\nu_s$
atmospheric neutrino oscillations may be checked in the future
in the Super-Kamiokande atmospheric neutrino experiment~\cite{sterile in atm}.
\item
$\nu_\mu\to\nu_\tau$
and
$\nu_e\to\nu_s$
oscillations
are strongly suppressed in SBL experiments.
With the approximation $c_s\simeq0$,
for the amplitude of
$\nu_\mu\to\nu_\tau$
oscillations
we have the upper bound
$ A_{\mu;\tau} \leq (a_e^0)^2 $,
that is shown in Fig.~\ref{fig7} (solid curve)
together with a recent exclusion curve obtained
in the CHORUS experiment (dash-dotted curve) and
the final sensitivity of the CHORUS and NOMAD experiments
(dash-dot-dotted curve)~\cite{CHORUS98}.
\end{itemize}
If these prediction will be falsified by future experiments
it could mean that
some of the indications
given by the results of neutrino oscillations experiments
are wrong and neither of the two four neutrino schemes A and B
is realized in nature,
or that Big-Bang Nucleosynthesis occurs with
a non-standard mechanism~\cite{foot}.

In conclusion,
we would like to emphasize that if
the analysis presented here is correct and one of the
two four neutrino schemes depicted in Fig.~\ref{fig8}
is realized in nature,
at the zeroth-order in the expansion over the small quantities
$c_e$ and $c_s$
the $4\times4$ neutrino mixing matrix has an extremely simple structure
in which the
$\nu_e,\nu_s$
and
$\nu_\mu,\nu_\tau$
sectors are decoupled:
in scheme A
\begin{equation}
U
\simeq
\left( \begin{array}{cccc}
0 & 0 & \cos\theta & \sin\theta \\
\cos\gamma & \sin\gamma & 0 & 0 \\
-\sin\gamma & \cos\gamma & 0 & 0 \\
0 & 0 & -\sin\theta & \cos\theta
\end{array} \right)
\label{mixA}
\end{equation}
and in scheme B
\begin{equation}
U
\simeq
\left( \begin{array}{cccc}
\cos\theta & \sin\theta & 0 & 0 \\
0 & 0 & \cos\gamma & \sin\gamma \\
0 & 0 & -\sin\gamma & \cos\gamma \\
-\sin\theta & \cos\theta & 0 & 0
\end{array} \right)
\,,
\label{mixB}
\end{equation}
where $\theta$ and $\gamma$ are, respectively,
the two-generation mixing angles relevant in
solar and atmospheric neutrino oscillations.
Therefore,
the oscillations of solar
and atmospheric neutrinos are independent
and
the two-generation analyses of solar and atmospheric neutrino oscillations
yield correct information on the mixing of four-neutrinos.

\bigskip

S.M.B. acknowledge
the support of the ``Sonderforschungsbereich 375-95 fuer
Astro-Teilchenphysik der Deutschen Forschungsgemeinschaf''.

\end{document}